# Comparative analysis of corneal and lens doses in nuclear medicine and impact of lead eyeglasses: a Monte Carlo simulation approach

Zahra Akbari Khanaposhtani[1], Hossein Rajabi*
*Department of Medical Physics, Tarbiat Modares University, Tehran, Iran*

*Corresponding Author email: hrajabi@modares.ac.ir

**Abstract**

Objective: Research on eye lens dosimetry for radiation workers has increased after the 2012 ICRP118 update on eye lens dose limits. However, corneal dosimetry remains underexplored due to historical focus and measurement challenges. This study uses a high-resolution digital eye phantom in Monte Carlo simulations to estimate corneal and lens doses for nuclear medicine staff, with and without lead glasses.

Method: The Monte Carlo code GATE (version 9.0) based on GEANT4 (version 10.6) was used to estimate and compare doses in a digital eye phantom, accounting for primary and scattered radiation from common radionuclides ($F^{18}$, $I^{131}$, $Tc^{99m}$) with varying lead glass shielding (0 to 0.75 mm).

Results: Across all radionuclides, the dose to the cornea was consistently higher than the dose to the lens. Notably, the ratio of corneal to lens dose increased with thicker lead glasses, indicating a greater dose reduction to the lens compared to the cornea.

Conclusion: The findings show that corneal doses from all studied radionuclides exceeded lens doses. Although increasing lead glass thickness reduced both, the reduction was more significant for the lens, raising the cornea-to-lens dose ratio. This trend suggests that while thicker lead glasses enhance lens protection, their practicality may be limited due to diminishing returns and potential discomfort.

## 1. Introduction

Radiation workers, including nuclear medicine personnel and interventional radiologists, are routinely exposed to ionizing radiation such as gamma rays and X-rays (Boice Jr J ,Dauer LT, Kase KR, Mettler Jr FA,Vetter RJ, 2020). Nuclear medicine professionals frequently work in close proximity to unsealed radionuclide sources, leading to localized radiation exposure. While the occupational exposure for nuclear medicine staff is primarily concentrated on the extremities, other organs and tissues generally receive minimal radiation exposure(Kollaard R, Zorz A, Dabin J, Covens P, Cooke J, Crabbé M, Cunha L, Dowling A, Ginjaume M, McNamara L., 2021). This distinction highlights the importance of targeted protective measures to mitigate localized radiation risks. Adherence to established safety protocols and advanced shielding techniques remain critical for minimizing occupational hazards in these high-risk fields.

The eye is a vital organ recognized for its sensitivity to ionizing radiation. According to the International Commission on Radiological Protection (ICRP), the lens of the eye is particularly vulnerable to radiation exposure (Hamada N, 2017; Smith, J., & Lee, A., 2023). Ionizing radiation is a physical agent that may lead to lens damage when exposure exceeds a certain threshold. Cataract formation, classified as a deterministic effect, typically occurs only above a well-established dose limit, below which the risk is considered negligible(Zhou, L., & Han, Y., 2022). Cataracts are classified as a deterministic effect of radiation exposure, characterized by a threshold dose below which the risk is minimal. Historically, the threshold for radiation-induced cataract formation was considered to be above 2 Gy(Ainsbury, Elizabeth A.; et al., 2009). However, emerging epidemiological and experimental evidence revealed that lens opacities can develop at lower doses than previously thought. This growing body of research led the International Commission on Radiological Protection (ICRP) to revise its recommendations in Publication 118(ICRP, 2012, p. 118), lowering the threshold to 0.5 Gy for chronic exposure. Accordingly, occupational dose limits for the lens of the eye were reduced to an average of 20 mSv per year over five years, with no single year exceeding 50 mSv. These changes reflect a heightened understanding of the eye's radiosensitivity and emphasize the need for stricter safety standards to protect radiation workers from long-term ocular damage. Understanding the susceptibility of the eye, particularly the lens, to ionizing radiation is critical for implementing protective measures and minimizing risks in environments where radiation exposure is prevalent. Ongoing research continues to refine our knowledge of dose thresholds and protective strategies, ensuring the preservation of ocular health in both occupational and medical settings.

Cataracts are a common eye condition that can lead to sensitivity to glare, halos around lights, difficulty with night vision, and altered color perception. While there is currently no cure for cataracts, their progression can be managed through updated eyeglass prescriptions or improved lighting conditions if they do not significantly impair vision. However, surgical intervention becomes necessary when cataracts begin to interfere with daily activities (Räsänen P, Krootila K, Sintonen H, Leivo T, Koivisto AM, Ryynänen OP, Blom M, Roine RP., 2006). Cataract surgery involves replacing the clouded natural lens with a synthetic intraocular lens, effectively restoring vision to a functional level in most cases. Despite the high success rate of this procedure, it is not without limitations. Research continues to explore pharmaceutical treatments as a potential alternative or complement to surgical intervention (Chen X, Xu J, Chen X, Yao K., 2021). These advancements promise more comprehensive and accessible management of cataracts in the future. In conclusion, while cataract surgery remains the gold standard for treatment, ongoing research into drug therapies may offer innovative solutions to address this prevalent condition.

The International Commission on Radiological Protection (ICRP) initially recommended an annual dose limit of 150 mSv for radiation workers (ICRP, 1990). However, This limit was subsequently reduced to an average of 20 mSv per year over five years, with no single year exceeding 50 mSv (ICRP, 2012). These regulatory limits are derived from threshold doses for deterministic effects such as cataract formation and reflect an evolving understanding of the eye's radiosensitivity. This revision demonstrates a commitment to prioritizing the safety and well-being of radiation workers (Hamada N, Azizova TV, Little MP., 2020). According to NCRP Commentary No. 26 (2016), the recommended equivalent dose limit for the lens of the eye is 50 mSv annually. This reflects the council's position at the time of publication and provides an official source for dose guidance in the United States(National Council on Radiation Protection and Measurements (NCRP), 2016). These updated guidelines reflect a global consensus on the importance of minimizing occupational radiation exposure and align with advancements in radiological science and protective measures. Such initiatives reflect the continued dedication of regulatory bodies to safeguarding individuals working in radiation-prone environments while promoting adherence to evidence-based safety standards.

As highlighted in NCRP Commentary No. 26(National Council on Radiation Protection and Measurements., 2022), it is essential to distinguish between absorbed dose and equivalent dose when assessing ocular radiation exposure. While absorbed dose (measured in Gray) reflects the energy deposited per unit mass, equivalent dose (measured in Sievert) accounts for the radiation type through a weighting factor. However, for deterministic effects such as cataracts, the absorbed dose is considered the more appropriate metric. This commentary also supports the ICRP's recommendations for using the operational quantity Hp(3) to better estimate the dose to the lens of the eye, particularly in occupational settings.

The International Commission on Radiological Protection (ICRP) Publication 118 has emphasized the importance of utilizing the operational quantity Hp (3) for surface dose measurements, particularly for monitoring the dose to the eye lens. Hp (3) is defined explicitly for assessing radiation exposure at a depth of 3 mm, making it highly relevant for evaluating the risks of cataract formation associated with ionizing radiation. This recommendation supports the need for precise and targeted monitoring to ensure compliance with safety standards and mitigate occupational health risks. Numerous studies have investigated occupational eye lens doses among radiation workers, comparing measured doses to established threshold levels. Notable research in this field includes contributions by Bellamy and Miodownik et al. (Bellamy MB, Miodownik D, Quinn B, Dauer L., 2020), Cornacchia and Errico et al. (Cornacchia S, Errico R, La Tegola L, Maldera A, Simeone G, Fusco V, Niccoli-Asabella A, Rubini G, Guglielmi G., 2019), and O'Connor and Walsh et al. (O′Connor U, Walsh C, Gallagher A, Dowling A, Guiney M, Ryan JM, McEniff N, O′Reilly G., 2015). These studies play a critical role in understanding radiation exposure risks and informing best practices for workplace safety. Adopting Hp (3) as an operational quantity reflects a commitment to advancing radiological protection measures, particularly in safeguarding the ocular health of individuals exposed to ionizing radiation in professional settings.

Numerous studies have examined occupational eye lens doses among healthcare workers, offering essential insights into exposure levels in clinical environments. For example, Bellamy et al. (Bellamy, M. B., Miodownik, D., Quinn, B., Dauer, L. T., n.d.) investigated interventional radiologists and cardiologists and reported that, despite protective measures, measured lens doses were often close to regulatory limits. Cornacchia et al. (Cornacchia, S., Errico, R., La Tegola, L., Maldera, A., Simeone, G., Fusco, V., Niccoli-Asabella, A., Rubini, G., Guglielmi, G., 2019) evaluated the lens dose in nuclear medicine and found that technologists working near unsealed sources are at elevated risk of exceeding annual dose constraints. Similarly, O'Connor et al. (O′Connor, U., Walsh, C., Gallagher, A., Dowling, A., Guiney, M., Ryan, J. M., McEniff, N., O′Reilly, G., 2015) highlighted dose variations among interventional radiology staff and emphasized the importance of using protective equipment such as lead glasses. These findings **demonstrate** the critical need for accurate monitoring, optimized shielding, and evidence-based dose reduction strategies in high-risk medical settings.

An analysis of ocular anatomy indicates that the cornea serves as the outermost tissue of the eye, making it the initial point of interaction with radiation. Gamma and X-ray radiation penetrate the cornea before reaching the lens, and **in some cases**, this may result in a higher radiation dose absorbed by the cornea than the lens. Being a thin structure, the cornea effectively absorbs low-energy photons (Volatier T, Schumacher B, Cursiefen C, Notara M., 2022). Additionally, secondary electrons generated by gamma and x-ray radiation interacting with the surrounding air can contribute to corneal radiation exposure (Kumar P, Watts C, Svimonishvili T, Gilmore M, Schamiloglu E., 2009; Nilsson BO, Brahme A., 1979). Similar interactions may occur with external factors such as eyeglasses or protective layers, potentially influencing the radiation dose received by the cornea. These findings highlight the importance of understanding surface-level radiation interactions with anterior ocular tissues, such as the cornea. While not associated with deterministic effects at low doses, corneal dose assessments may offer complementary insight into exposure distribution, particularly in high-risk or high-frequency clinical environments.

With regards to corneal injuries, it is important to expand on the relationship between ionizing radiation and its potential to damage the cornea. Although keratitis is primarily associated with UV radiation, ionizing radiation has also been reported as a contributing factor in rare but significant cases(Ting D. S. J. et al., 2021).The cornea, being the first ocular tissue to interact with radiation, is particularly susceptible to injury from low-energy photons and secondary electrons(Nuzzi R, Trossarello M, Bartoncini S, Marolo P, Franco P, Mantovani C, Ricardi U., 2020). Despite this vulnerability, most protective and dosimetric efforts have historically centered on the lens. This **points to** the importance of establishing a clear rationale for including corneal dose assessments in occupational radiation safety practices(Ciraj-Bjelac & Rehani, 2014).

Corneal injuries are often more acute and associated with severe complications compared to lens damage, frequently resulting in significant and potentially irreversible vision impairment. Keratitis, an inflammation of the cornea, remains a leading cause of corneal blindness globally(Rahmani, S., Javadi, M. A., Hassanpour, K., & Karimian, F., 2022). However, these effects are primarily associated with external or infectious causes rather than ionizing radiation. Radiation-related corneal injury is uncommon and typically limited to high-dose therapeutic exposures. Severe damage to corneal tissues may require advanced medical interventions, including surgical procedures such as corneal transplantation or keratoplasty (Herring IP., 2003). Additionally, the impact of radiation therapy on ocular structures, including the cornea, particularly in therapeutic applications such as proton therapy and brachytherapy, is a subject of ongoing research. While the broader effects of radiation on the ocular system are well-documented (Nuzzi R, Trossarello M, Bartoncini S, Marolo P, Franco P, Mantovani C, Ricardi U., 2020), recent studies have further explored its specific influence on corneal health. A noteworthy publication by Zahariev, Hristov, et al. (V. Zahariev, N. Hristov, St. Vizev, P. Angelova., 2024) provides new insights into the implications of radiation exposure on the eye, emphasizing the importance of preventive measures and targeted therapeutic approaches. Given the critical nature of corneal health in maintaining vision, a comprehensive understanding of injury mechanisms and treatment modalities is essential for mitigating long-term ocular damage and preserving patient quality of life.

To evaluate the impact of lead eye protection glasses on the absorbed dose to the cornea and lens, we assessed the radiation dose corresponding to varying thicknesses of protective eyewear placed in front of an eye phantom. Subsequently, the absorbed dose distribution within the corneal and lens regions of the eye phantom was calculated.

With updated radiation safety regulations and growing concerns about cataracts at low doses (ICRP Publication 118)(ICRP, 2012, p. 118), significant attention has been directed toward eye dosimetry for radiation-exposed personnel. This has led to numerous experimental and simulation-based investigations focusing on this critical area. Experimental studies, such as those by Wrzesień & Królicki (Wrzesień M, Królicki L, Olszewski J., 2018) and Martinez & Gupta (Martinez, R., Gupta, N., Alvarez, T., & Chen, L., 2021), have provided valuable empirical data. Complementarily, advanced Monte Carlo simulations, as demonstrated by Morhrasi & Jumpee (Morhrasi P, Jumpee C, Charoenphun P, Chuamsaamarkkee K., 2022)and Hoeijmakers & Hoenen(Hoeijmakers EJ, Hoenen

K, Bauwens M, Eekers DB, Jeukens CR, Wierts R., 2024), have offered detailed insights into dose distributions and protective measures.

Recognizing the cornea as a sensitive tissue alongside the lens is of dosimetric interest, particularly in contexts where radiation may impact multiple ocular structures. While the lens remains the primary concern in occupational exposure, evaluating corneal dose distributions using Monte Carlo simulations can offer additional insight into overall eye dose assessments. This study employed such simulations to estimate radiation doses to both the lens and cornea, contributing to a more comprehensive understanding of ocular dosimetry.

The simulations incorporated various photon beams emitted by radionuclides commonly used in nuclear medicine and X-rays utilized in interventional radiology. We modeled scenarios where staff members wore lead glasses of varying thicknesses to assess protective measures. The resulting data allowed for a comparative analysis of radiation exposure to the cornea and lens under these protective conditions.

Our findings underscore the necessity of considering corneal exposure in occupational radiation safety protocols. This research highlights the potential for enhanced protective strategies to mitigate risks to sensitive ocular tissues, thereby contributing to the overall safety of medical professionals in radiation-intensive environments.

## 2. Methods and Materials

### 2.1. *Monte Carlo simulation*

GATE (version 9.0) was employed to calculate the absorbed doses corresponding to the specific geometry utilized in this study (Strulab D, Santin G, Lazaro D, Breton V, Morel C., 2003)**.** GATE is a Monte Carlo simulation toolkit developed based on GEANT4 (version 10.6) and tailored explicitly for nuclear medicine applications. This advanced simulation framework accounts for all potential photon and electron interactions, including photoelectric absorption, coherent and incoherent scattering, ionization, x-ray production, and Auger electron emission, during particle tracking within materials.

The simulation framework offers four distinct physics models, with the selection dependent on the specific application. Among the available physics models, the **Livermore physics model** was selected for its capability to simulate electromagnetic interactions at low energies, which is essential for accurate ocular dosimetry. Specifically, this model allows particle tracking down to a **threshold of 250 eV**, ensuring that low-energy photons and electrons—which significantly contribute to eye dose—are accurately simulated. This threshold is commonly applied in radiological simulations for the eye, based on its suitability for modeling tissue-relevant low-energy interactions (Chauvie, S., et al., 2007; Han, J. et al., 2020). The Livermore model is based on evaluated data libraries (EPDL97, EEDL, and EADL) and provides accurate modeling of electromagnetic interactions for low-energy photons and electrons(Chauvie, S., et al., 2007). Any remaining particle energy below this threshold is absorbed locally.

The simulation employed a step size of 0.01 mm for all particles, ensuring high spatial resolution. To achieve robust statistical accuracy, the number of histories was set at 200 million, resulting in dose uncertainties in the cornea and lens of the eye phantom being maintained below 0.01. The calculated doses were then normalized per MBq·s (i.e., 1 MBq for one second), which corresponds to one million decays, aligning the simulation output with clinical activity-based dose references. This setup ensured precise and reliable dosimetric evaluations critical to the study's objectives.

### *2.2. Geometry of Simulation*

This study acquired a three-dimensional (3D) ocular model in OBJ format developed by Wavefront Technologies (Mello GR, Rocha KM, Santhiago MR, Smadja D, Krueger RR., 2012). The model, representing a half-eye for educational purposes, was processed using the Binvox open-source software to create a voxelized 3D phantom, as illustrated in Figure 1 (Abdrakhmanov Renat,Kerimbek Imangali, 2024). The phantom's resolution was set at 0.039×0.046×0.097 mm. While the model included key ocular structures, only the cornea, lens, and associated components were utilized. MATLAB codes these structures for distinction during simulation (Gilat A, 2017). The atomic composition of the tissues was derived from ICRP 89 (ICRP, 2002), detailed in Table 1. Subsequently, the phantom was converted into Interfile format using XMedCon software (Liao W, Deserno TM, Spitzer K., 2008), compatible with the Monte Carlo toolkit (Sarrut D, Baudier T, Borys D, Etxebeste A, Fuchs H, Gajewski J, Grevillot L, Jan S, Kagadis GC, Kang HG, Kirov A., 2022). Protective lead glasses were modeled as a box (30×30×X mm) positioned 9 mm from the eye. This distance was selected to reflect a typical anatomical gap between the eye and eyewear in clinical settings. Additionally, the Monte Carlo simulation accounted for forward scattering of secondary electrons, including photoelectric and Compton electrons, ensuring a more comprehensive dosimetric assessment. Lead thicknesses ranged from 0 to 0.75 mm (Hirata Y, Fujibuchi T, Fujita K, Igarashi T, Nishimaru E, Horita S, Sakurai R, Ono K., 2019), with a density of 11.4 g/cm³. This density corresponds to pure elemental lead, which is often used as a reference in simulation studies. However, in real-world applications, lead glasses may incorporate lead-based compounds such as lead oxide (PbO), which have slightly different effective densities depending on the formulation and manufacturing process(Fujibuchi, T. et al., 2019). While this study used a standard lead density to ensure consistency and comparability with previous dosimetric models, future work may explore material-specific simulations to account for variations in commercial shielding products. Zero thickness indicated the absence of protection. A schematic representation of the geometry is provided in Figure 1.

While our voxelized eye model captures key anatomical structures relevant to radiation dose assessment—such as the cornea and lens—it is relatively simplified compared to more anatomically detailed models, such as the one described by Behrens et al. (Behrens, R.; Dietze, G.; Zankl, M., 2009). The Behrens model includes more precise curvature and layered tissue segmentation of the eye. In contrast, our model emphasizes structural clarity and computational feasibility, which are particularly advantageous for large-scale Monte Carlo simulations involving multiple dose scenarios and shielding conditions. Despite this simplification, the corneal and lens volumes in our model are within accepted anatomical ranges, supporting valid comparative dose analysis. Future studies may incorporate more complex geometries to refine spatial resolution and enhance dosimetric accuracy for smaller ocular substructures.

**Table 1**
**Atomic composition (by weight fraction) of different tissues of the eye used in the simulation.**

| Tissue | Weight Fraction (%) | | | | | | | | | |
|---|---|---|---|---|---|---|---|---|---|---|
| | H | C | N | O | Na | P | S | Cl | K | Ar |
| Lens | 0.096 | 0.195 | 0.057 | 0.646 | 0.001 | 0.001 | 0.003 | 0.001 | - | - |
| Eyeball | 0.107 | 0.069 | 0.017 | 0.803 | - | 0.001 | 0.001 | - | - | 0.002 |
| cornea* | 0.102 | 0.143 | 0.034 | 0.708 | 0.002 | 0.003 | 0.003 | 0.002 | 0.003 | - |

* The ICRU-recommended eye soft tissue was used as a proxy for the cornea tissue.

**Figure 1**
**The schematic representation of geometry used in this study. The thickness of eyeglass changes from zero to 0.75 mm. Since radiation was considered orthogonal to the eyeglass, As the distance increased, the protective effect of the lead glasses diminished, effectively reducing the shielding benefit to negligible levels.**

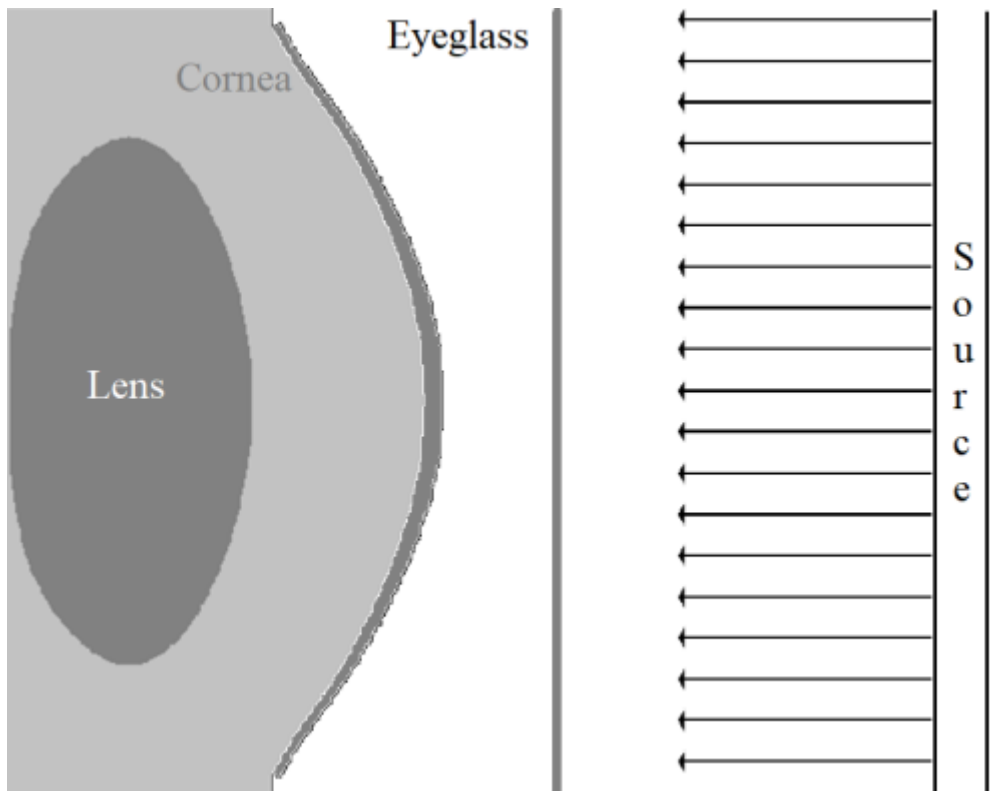

## *2.3. Spectra of Radiation*

In this study, three commonly used radionuclides in nuclear medicine—fluorine-18 (F-18), technetium-99m (Tc-99m), and iodine-131 (I-131)—were selected to evaluate the absorbed dose to the cornea and lens. These isotopes represent a range of photon energies (140 keV for Tc-99m, 511 keV for F-18, and 364 keV for I-131), enabling comprehensive assessment across diagnostic and therapeutic energy levels (Kearfott, K. J., Rucker, R. H., Samei, E., & Mahesh, M., 2001; Stabin, M. G., & Siegel, J. A., 2003). The energy spectra of radionuclides $I^{131}$, $Tc^{99m}$, and $F^{18}$ were obtained from the reference text "The MIRD Radionuclides and Decay Scheme" **.** For this study, only the x-ray and gamma radiation spectra were utilized, as other forms of radiation are negligible at a distance from the radionuclide source. These spectra simulated radiation exposure in a controlled hot laboratory environment.

Additionally, two x-ray spectra with peak energies of 30 and 50 keV were included to represent scattered radiation typically encountered in interventional radiology procedures. These energy levels were chosen based on published scatter energy distributions, **reflecting the most probable photon energies reaching the eye level during fluoroscopic procedures, particularly in cardiac and other interventional radiology applications where soft x-rays are prevalent and lead eye goggles demonstrate their greatest protective effectiveness**(Miller, D. L., Vano, E., Bartal, G., Balter, S., Dixon, R. G., Padovani, R., ... & Niklason, L. T., 2010; Vano, E., Gonzalez, L., Fernandez, J. M., Haskal, Z. J., & Rehani, M. M., 2008).

This addition allows for comparison between radionuclide-based exposure in nuclear medicine and secondary radiation exposures from x-rays used in interventional settings. The selected spectra help explore how different photon energies interact with ocular tissues and protective eyewear, providing valuable insight into energy-dependent shielding effectiveness and biological dose deposition (Seibert, J. A., Boone, J. M., & Gagne, R. M., 2004). This experimental setup allowed for the precise evaluation of radiation interactions while maintaining standardized conditions**.** However, in real-life clinical settings, technologists often observe the source at a downward angle. This deviation in head orientation could influence the radiation dose received by the eye and may lead to variation in shielding effectiveness. Further investigations incorporating angular incidence would enhance the realism and applicability of simulation findings(Vanhavere, F., Carinou, E., Gualdrini, G., Clairand, I., & Denozière, M., 2008).

### *2.4. Estimation of the corneal and lens doses*

The energy spectra of radiation generated in the preceding stages were utilized to irradiate an eye phantom. This radiation was applied uniformly and parallelly across the entire phantom, with the source designed to encompass the whole cross-sectional area of the eye model. Each simulation generated two binary output files of dimensions comparable to the eye phantom. The first file contained voxel-specific dose data, while the second provided the associated uncertainties for each voxel.

As Gilat (Gilat A, 2017) described, a custom MATLAB script was employed to extract dose values and their corresponding uncertainties for the phantom tissues. Despite the large number of voxels analyzed, the maximum statistical uncertainty recorded across all simulation outputs was determined to be 0.01 (1%), which indicates that the relative error in the calculated absorbed dose per voxel is within 1%. In Monte Carlo simulations, this metric reflects the confidence in dose estimation due to the finite number of particle histories used, ensuring high precision and reliability of results.

The values in Table 2 represent dose conversion factors, expressed in mGy per one million incident photons (i.e., per 1 MBq·s assuming one million decays). These values do not reflect absolute clinical dose, but instead provide a standardized metric for comparing dose deposition across different scenarios. To express the dose in μGy/MBq, we normalized the results to one disintegration per photon, consistent with the assumption of 1 MBq·s producing one million decays. Although the simulation setup assumes parallel and orthogonal beam geometry, real-life exposure may involve varying incident angles depending on technologist position and source geometry. As such, our results represent idealized baseline conditions, which could be expanded upon in future work using more anatomically dynamic models and angular distributions.

## 3. Result

### *3.1. Dose to different parts of eye:*

Table 2 comprehensively summarizes the radiation dose to various eye components, including the lens, cornea, and total ocular tissues. A pooled dataset encompassing all radionuclides was analyzed using a paired t-test to evaluate dose distribution. The analysis revealed no statistically significant differences between the dose to the lens and the overall ocular tissues ($p$-value $< 0.001$). Conversely, a significant difference was observed between the dose to the cornea and other ocular tissues ($p$-value $> 0.75$).

These findings indicate that the radiation dose to the lens can be considered a reliable representation of the total ocular dose. This conclusion has important implications for simplifying dose assessment protocols in radiological studies, as it underscores the lens as an appropriate surrogate for evaluating overall ocular exposure**.**

While absolute dose values for radionuclides are provided in Table 2, X-ray simulations were used specifically to evaluate cornea-to-lens dose ratios (Table 3), and thus their results are not included here due to methodological differences.

**Table 2**

**The doses to the lens, cornea and the eye tissues in mGy per one million photons from the radionuclides.**

| *Eye glass* | *$F^{18}$* | | | *$I^{131}$* | | | *$Tc^{99m}$* | | |
|---|---|---|---|---|---|---|---|---|---|
| *(mm)* | lens | cornea | eye | lens | cornea | eye | lens | cornea | eye |
| *0.0* | 8.21 | 9.06 | 8.39 | 5.69 | 6.23 | 5.82 | 1.75 | 2.07 | 1.82 |
| *0.05* | 8.13 | 9.01 | 8.32 | 5.62 | 6.16 | 5.74 | 1.53 | 1.83 | 1.60 |
| *0.07* | 8.11 | 8.96 | 8.30 | 5.59 | 6.13 | 5.71 | 1.45 | 1.73 | 1.52 |
| *0.10* | 8.07 | 8.93 | 8.26 | 5.54 | 6.08 | 5.66 | 1.34 | 1.61 | 1.41 |
| *0.20* | 7.95 | 8.79 | 8.13 | 5.38 | 5.92 | 5.51 | 1.04 | 1.24 | 1.08 |
| *0.30* | 7.82 | 8.66 | 8.00 | 5.24 | 5.75 | 5.36 | 0.80 | 0.96 | 0.83 |
| *0.40* | 7.69 | 8.53 | 7.88 | 5.11 | 5.62 | 5.22 | 0.61 | 0.74 | 0.64 |
| *0.50* | 7.57 | 8.45 | 7.76 | 4.97 | 5.48 | 5.08 | 0.47 | 0.57 | 0.50 |
| *0.60* | 7.45 | 8.31 | 7.64 | 4.84 | 5.34 | 4.95 | 0.36 | 0.44 | 0.38 |
| *0.70* | 7.33 | 8.18 | 7.52 | 4.71 | 5.22 | 4.82 | 0.28 | 0.34 | 0.29 |
| *0.75* | 7.27 | 8.10 | 7.46 | 4.64 | 5.14 | 4.76 | 0.24 | 0.30 | 0.26 |

### *3.3. The ratio of cornea to lens dose*

Table 3 presents the ratio of cornea-to-lens radiation exposure, considering photon energy levels and the thickness of lead glasses used for eye protection. The results indicate an inverse relationship between this ratio and photon energy. Specifically, as the mean photon energy decreases, the cornea-to-lens ratio increases. Understanding the cornea-to-lens dose ratio is important because it reflects how radiation interacts with anterior ocular tissues and helps assess whether shielding strategies designed for the lens also offer protection for the cornea.

The mean photon energies for $F^{18}$, $I^{131}$, and $Tc^{99m}$ radionuclides are approximately 511 keV, 364 keV, and 140 keV, respectively (IAEA, 2000).These values represent the dominant gamma or annihilation photon emissions used in diagnostic or therapeutic procedures. Mean photon energy is used here as a metric to characterize the penetrating ability and interaction probability of each radionuclide with ocular tissues. Higher-energy photons tend to deposit dose deeper in tissue, influencing the dose distribution between the cornea and lens.

Similarly, the cornea-to-lens ratio for beta radiation follows a comparable trend: beta emitters like Tc-99m or Lu-177 deliver relatively **higher dose to the cornea than to the lens**, as indicated by dose conversion coefficients from recent Monte Carlo and validated experimental work(Hoeijmakers EJ, Hoenen K, Bauwens M, Eekers DB, Jeukens CR, Wierts R., 2024)

Furthermore, a positive correlation is observed between the cornea-to-lens ratio and the thickness of lead glasses. Increasing the thickness of lead shielding reduces the absorbed dose to both the cornea and lens, even as the average radiation energy decreases. For monoenergetic x-ray electrons at 30 keV, the cornea-to-lens ratio is approximately 1.8 times that at 50 keV without shielding. However, as the thickness of lead glass increases, this ratio diminishes significantly. These findings highlight the importance of optimizing lead glass shielding to mitigate radiation exposure effectively while considering photon energy levels.

The X-ray simulations in this study utilized monochromatic photon beams (30 keV and 50 keV) as a simplified approach to explore cornea-to-lens dose ratios. While such simplification aids in isolating the influence of photon energy on dose distribution, it does not fully represent the spectral characteristics of diagnostic or interventional X-ray sources. In clinical practice, occupational eye exposure is predominantly due to scattered radiation with broad spectral distributions, not direct primary beams. Therefore, these results may underestimate the actual protection factors provided by lead eyewear. Future studies will incorporate realistic spectra and scattering conditions to better reflect clinical exposure scenarios.

**Table 3**
**The ratio of cornea-to-lens dose for all types of radiation examined in this study.**

| | *Radiation* | | | | |
|---|---|---|---|---|---|
| *Eye glass (mm)* | $F^{18}$ | $I^{131}$ | $Tc^{99m}$ | X-ray 50 keV | X-ray 30 keV |
| *0.0* | 1.104 | 1.094 | 1.187 | 3.81 | 6.80 |
| *0.05* | 1.107 | 1.098 | 1.191 | 1.60 | 2.72 |
| *0.07* | 1.105 | 1.097 | 1.191 | - | - |
| *0.10* | 1.106 | 1.097 | 1.195 | 1.47 | 1.93 |
| *0.20* | 1.105 | 1.099 | 1.199 | 1.44 | 1.70 |
| *0.30* | 1.108 | 1.097 | 1.200 | | |
| *0.40* | 1.109 | 1.102 | 1.206 | | |
| *0.50* | 1.116 | 1.102 | 1.208 | | |
| *0.60* | 1.115 | 1.105 | 1.216 | | |
| *0.70* | 1.116 | 1.108 | 1.220 | | |
| *0.75* | 1.114 | 1.108 | 1.223 | | |

## *3.4. Protection efficiency of the lead eyeglasses*

Figure 2 illustrates the protection efficiency of lead glasses when exposed to various radionuclides investigated in this study. Protection efficiency is defined as the percentage reduction in radiation dose achieved by wearing the lead glasses compared to no shielding. The results indicate that the protective capability of lead glasses for nuclear medicine personnel is limited. However, this limitation is influenced by the energy of the radiation emitted from different radionuclides. For instance, lead glasses offer greater attenuation for lower-energy photons (e.g., $Tc^{99m}$ at 140 keV) but become less effective for higher-energy emissions like $F^{18}$ at 511 keV. Therefore, the degree of protection varies depending on the isotope in use and the corresponding photon energy levels (Wagner, L. K., Eifel, P. J., & Geise, R. A., 1994). The thickness values simulated in this study refer to lead-equivalent thicknesses, which are used to represent the radiation attenuation capacity of different protective eyewear materials. This standardization enables more accurate comparisons across shielding products regardless of their composition or actual physical thickness.

For glasses with a thickness of 0.75 mm, the protection efficiency was 11% for $F^{18}$ radiation and 18% for $I^{131}$ radiation. These values suggest minimal shielding effectiveness for these radionuclides. However, a significant improvement in protection was noted for $Tc^{99m}$ radiation, where the efficiency exceeded 85%. It demonstrates that lead glasses provide substantial shielding against $Tc^{99m}$ but are less effective for other radionuclides tested.

These findings highlight the importance of selecting appropriate protective equipment based on the specific radionuclides encountered in clinical or laboratory settings to ensure optimal safety for nuclear medicine professionals. Further investigation into enhancing the protective properties of lead glasses may be warranted to address the limitations observed for specific radionuclides.

To evaluate the effectiveness of lead glasses, a protection efficiency exceeding 50% was considered clinically beneficial based on prior shielding benchmarks in diagnostic imaging. This threshold aligns with the ALARA (As Low As Reasonably Achievable) principle, which emphasizes minimizing radiation exposure while balancing the practicality and comfort of protective measures. Although lead glasses provide significant dose reduction for low-energy photons such as Tc-99m, they offer limited protection for high-energy emissions like F-18. Additionally, potential drawbacks such as discomfort, reduced visibility, and user compliance should be considered when weighing the cost-benefit of wearing such equipment in clinical settings (Vano, E., Gonzalez, L.,

Fernandez, J. M., & Haskal, Z. J., 2008).

**Figure 2**

**Protection efficiency of lead eyeglasses with varying lead-equivalent thicknesses for different radionuclides. Monte Carlo simulations using the Livermore model were conducted to evaluate dose reduction to the eye. Error bars represent statistical uncertainties, all maintained below 1%, and may not be visible due to their small magnitude. This type of glass has considerable efficiency for $Tc^{99m}$ radiation. However, they are not protective for the radiation from $F^{18}$ and $I^{131}$ radiations.**

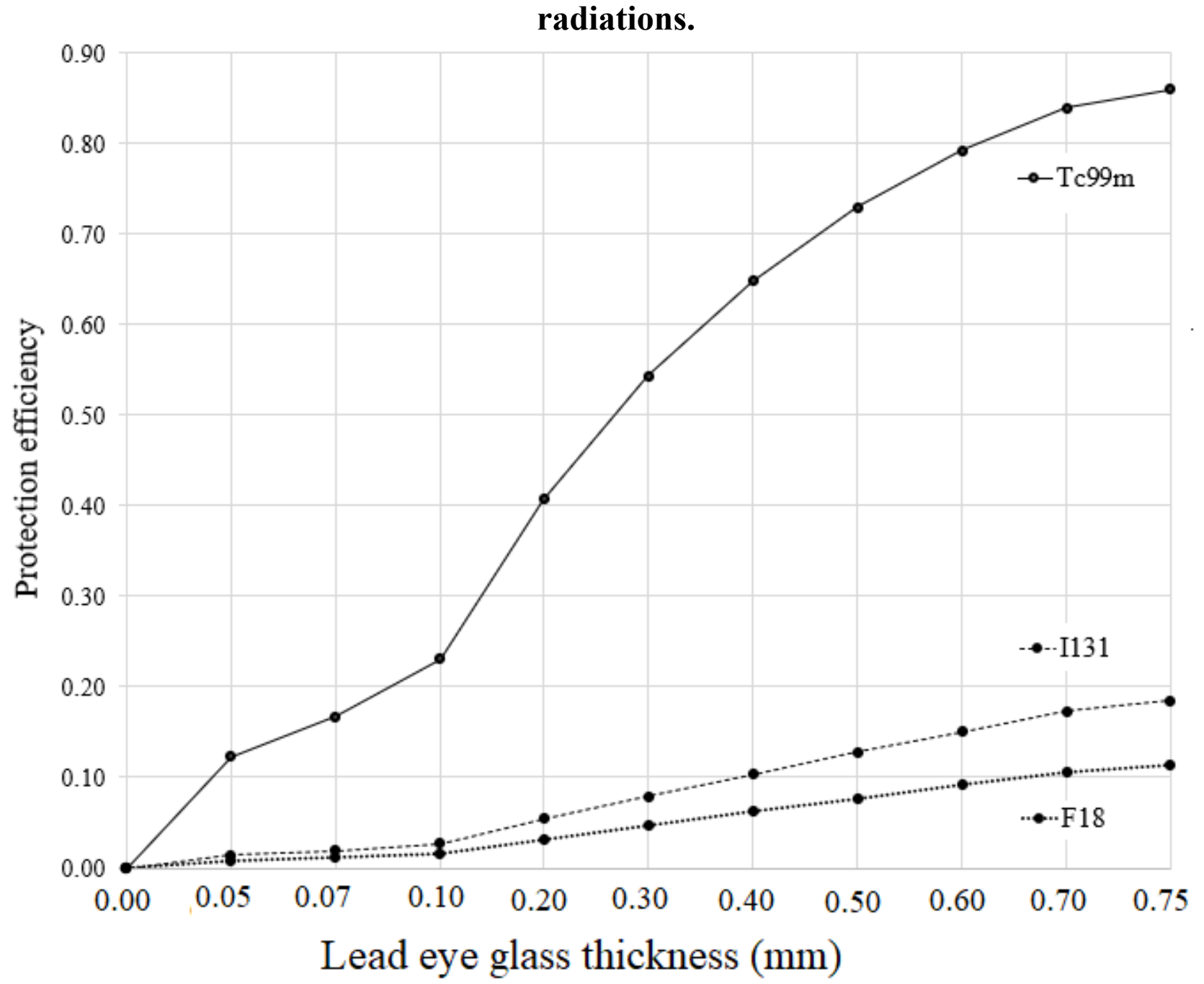

## 4. Discussion

Recent studies have increasingly focused on occupational eye dosimetry due to updated radiation protection guidelines and the recognized risk of cataract formation from low-dose exposures. Our findings contribute to this ongoing effort by evaluating the effectiveness of lead eyewear in reducing radiation dose to the eye

These findings underscore the importance of evaluating protective eyewear's effectiveness in occupational settings, ensuring compliance with evolving safety standards while mitigating risks of radiation-induced ocular damage.

An experimental study conducted by Guiu-Souto and Sánchez-García (Guiu-Souto J, Sánchez-García M, Vázquez-Vázquez R, Otero C, Luna V, Mosquera J, Busto RL, Aguiar P, Ruibal A, Pardo-Montero J, Pombar-Camean M., 2016) utilized silicon dosimeters to quantify occupational lens doses in PET centers. Their findings revealed that 59% and 40% of the total dose were attributed to the preparation of fluorodeoxyglucose (FDG) syringes and the injection process, respectively. It implies that all other operational tasks contribute only 1% of the total dose received by staff.

Further research has highlighted discrepancies between experimental studies and computational simulations. For instance, da Silva et al.(da Silva EH, Martin CJ, Vanhavere F, Buls N., 2020) reported lower dose estimations in experimental studies compared to Monte Carlo simulations, suggesting that simulations may overestimate exposure under certain conditions. This discrepancy is often attributed to the use of simplified geometries and low-resolution phantoms in simulations, as noted by Morhrasi et al. (and Rajkhan et al. (Morhrasi P, Jumpee C, Charoenphun P, Chuamsaamarkkee K., 2022) (Rajkhan A, Alsafi K, Banoqitah E, Subahi A., 2019) In contrast, our study utilized a high-resolution voxel phantom, which may offer a more accurate anatomical representation and contribute to more precise dose estimations. This methodological improvement could explain some of the differences between our findings and previous experimental data. Overall, these comparisons reinforce the importance of phantom resolution in dose assessment and support the reliability of our simulation approach**.**

These findings indicate the need for advanced simulation techniques to enhance dose estimation accuracy, thereby improving radiation safety protocols for healthcare professionals in PET centers.

Miyaji et al. (Miyaji N, Miwa K, Iimori T, Wagatsuma K, Tsushima H, Yokotsuka N, Murata T, Kasahara T, Terauchi T., 2022) investigated occupational eye lens doses in nuclear medicine in a study closely related to our work. They utilized a RANDO human phantom equipped with dosimeters to experimentally estimate lens doses. Their methodology involved radionuclides diluted in 3 mL water within a 5 mL syringe as the radiation source. Reported doses were 3.30 μGy/MBq for $F^{18}$ and 1.91 μGy/MBq for $Tc^{99m}$.

Similarly, Rajkhan et al. (Rajkhan A, Alsafi K, Banoqitah E, Subahi A., 2019) conducted a simulation study employing the MCNP5 Monte Carlo code to estimate eye doses from radioactive solutions in syringes. Their findings indicated dose rates of 20.55, 7.90, and 2.33 μSv/MBq for $F^{18}$, $I^{131}$, and $Tc^{99m}$, respectively.

McCann et al. (McCann A, Cournane S, Dowling A, Maguire D, Lucey J, Leon L., 2021) further explored eye dosimetry using the EGSnrc Monte Carlo code through experimental and simulation methods. They assessed dose rates from unshielded and shielded syringes containing $Ga^{68}$ and $F^{18}$, reporting approximately threefold higher doses for $Ga^{68}$ than $F^{18}$, though no absolute values were provided.

Our study demonstrated doses of 8.21 μGy/MBq for F18, 1.75 μGy/MBq for Tc99m, and 5.69 μGy/MBq for I131, which align well with previous empirical studies and corroborate their findings(da Silva EH, Martin CJ, Vanhavere F, Buls N., 2020; Guiu-Souto J, Sánchez-García M, Vázquez-Vázquez R, Otero C, Luna V, Mosquera J, Busto RL, Aguiar P, Ruibal A, Pardo-Montero J, Pombar-Camean M., 2016; Martinez, R., Gupta, N., Alvarez, T., & Chen, L., 2021; Miyaji N, Miwa K, Iimori T, Wagatsuma K, Tsushima H, Yokotsuka N, Murata T, Kasahara T, Terauchi T., 2022; Rajkhan A, Alsafi K, Banoqitah E, Subahi A., 2019). Here, MBq refers to one megabecquerel for a duration of one second (MBq·s), corresponding to one million disintegrations, which was used for dose normalization in the simulation. Although $F^{18}$ exhibited the highest dose per unit activity, the larger activities typically administered for $I^{131}$ treatments in clinical practice may result in higher total exposure to technologists from $I^{131}$ overall. In this simulation, dose values were calculated **per megabecquerel (MBq)** for a fixed geometry, assuming a standard source-to-eye distance and a constant exposure duration. This normalization allowed for direct comparison between radionuclides independent of clinical variations in time or positioning. While time and distance are critical in real-world settings, our approach focuses on the **relative dose potential per unit activity**. Actual occupational exposure would further depend on factors such as procedure duration, shielding practices, and handling frequency, which were not simulated here but are acknowledged as influential.

In Table 2, the radiation doses to various parts of the eye are presented. The data indicate that $F^{18}$ delivers a significantly higher dose than other radionuclides. Specifically, the dose from $F^{18}$ was observed to be 1.44 times higher than $I^{131}$ and 4.6 times higher than $Tc^{99m}$. Analysis of the cumulative data for all radionuclides revealed no statistically significant differences between the lens dose and the total eye dose. This finding supports the conclusion that the lens dose can serve as a reliable representation of the total eye dose, which is crucial for ensuring the safety of radiation workers.

Conversely, significant differences were observed between the corneal and total eye doses. Notably, the corneal dose was approximately 10% higher than the lens dose for $F^{18}$ and $I^{131}$, while for $Tc^{99m}$, this difference increased to approximately 19% (as shown in Table 3). These findings reveal the need to carefully consider corneal exposure in radiation protection protocols, mainly when using radionuclides with higher relative corneal doses.

The relationship between the cornea-to-lens dose ratio and the thickness of lead glasses reveals notable trends. As the thickness of lead glasses increases, the absorbed dose to both the cornea and lens decreases. However, the cornea-to-lens dose ratio rises with increasing eye shield thickness, indicating greater shielding effectiveness for the lens than the cornea. Specifically, the lens demonstrates enhanced protection with effectiveness increments of 0.010, 0.014, and 0.036 for $F^{18}$, $I^{131}$, and $Tc^{99m}$, respectively.

This phenomenon may be attributed to the sequential passage of radiation, which first traverses the cornea before reaching the lens. Additionally, electrons generated near the corneal surface contribute to this disparity. For instance, as shown in Table 3, 30 keV electrons impart significantly higher doses to the cornea than 50 keV. Furthermore, bremsstrahlung radiation produced by lead shielding can amplify this effect.

While lead glasses mitigate overall radiation exposure, their impact on dose distribution requires careful consideration. These findings highlight the need for optimized shielding designs to effectively balance protection for both ocular components.

In clinical practice, ceiling-suspended shields or bench-mounted lead glass barriers are frequently employed during the preparation and administration of radiopharmaceuticals. These fixed shielding devices are particularly useful for high-energy radionuclides such as F18, providing effective eye dose reduction while improving comfort and user compliance compared to lead eyewear. Their integration into routine workflows offers a practical alternative or complement to personal protective equipment, especially in procedures involving prolonged exposure or repeated handling of radioactive materials.

In addition to the angular positioning of the eye relative to the radiation source, additional safety measures can significantly reduce radiation exposure. For instance, the use of **L-block shields**—which are lead glass barriers placed between the radioactive source and the worker—has been shown to effectively attenuate scattered radiation during radiopharmaceutical preparation and injection, particularly in PET settings(IAEA, 2004).

Moreover, the increasing adoption of **automated injectors** in PET facilities has helped minimize direct handling of radiotracers, leading to significant reductions in extremity and eye doses for technologists (Bebbington, N.A., et al., 2015; Fahey, F.H., 2011). These systems automate the syringe preparation and injection process, thus increasing operational safety and efficiency.

Based on this analysis, standard protective eyewear does not provide adequate protection against $F^{18}$ and $I^{131}$ radiation exposure. Consequently, leaded glass materials are strongly recommended to enhance ocular safety for personnel working with these specific radionuclides. It calls attention to the need for tailored protective measures in radiation safety protocols to address varying exposure risk levels associated with radionuclides.

## 5. Conclusion

Our study revealed that the absorbed dose to the cornea consistently exceeded that to the lens across all radionuclides evaluated. While this highlights the importance of including corneal dosimetry—alongside lens dosimetry—for a comprehensive assessment of ocular radiation exposure, it is important to note that the corneal dose levels observed in typical occupational settings are unlikely to cause biological harm. However, corneal dose tracking may still be useful in the context of prolonged exposure, shielding optimization, and safety protocol development.

While lead glasses reduce overall radiation exposure, they may increase the **cornea-to-lens dose ratio**, indicating unequal shielding effectiveness. F18 and I131 in particular demonstrated **low shielding efficiency**, whereas Tc99m showed considerable dose reduction. These results point to a critical need for radionuclide-specific shielding strategies.

Ultimately, our findings advocate for **tailored eye protection**, especially in high-risk scenarios involving F18 or I131, and call for further innovation in the design of shielding devices. Incorporating both lens and cornea protection will be essential in optimizing safety standards for nuclear medicine personnel.

Moreover, while leaded eyewear provides a measurable reduction in eye dose, particularly for low-energy photons, it is essential to consider its potential impact on work efficiency. Studies have reported that cumbersome or uncomfortable PPE may reduce task performance or increase procedural time, especially in high-intensity clinical settings(Dauer, Lawrence T; Miller, Donald L; Schueler, Beth A; Damilakis, John; Vano, Eliseo; Rehani, Madan M, 2013). If the time spent performing radiation-related tasks increases significantly due to PPE usage, the cumulative exposure might inadvertently rise. For instance, in scenarios involving high-energy radionuclides where dose reduction from leaded glasses is approximately 10%, any workflow slowdown exceeding this threshold could result in a net increase in eye dose.

According to ICRP recommendations, the annual equivalent dose limit of 150 mSv applies specifically to the lens of the eye for radiation workers, whereas the whole-body effective dose limit is 20 mSv. Clarifying this distinction is critical when evaluating occupational radiation protection.

Thus, optimal PPE design must balance protective benefits with ergonomic efficiency to ensure overall dose reduction.

## References

Abdrakhmanov Renat,Kerimbek Imangali. (2024). Learning Latent Representations for 3D Voxel Grid Generation using Variational Autoencoders. *IEEE AITU: Digital Generation*, 169–173. https://doi.org/10.1109/IEEECONF61558.2024.10585546

Ainsbury, Elizabeth A.;, Bouffler, Simon D.;, Dorr, Wolfgang;, Graw, Jochen;, Muirhead, Colin R.;, Edwards, Anthony A.;, & Cooper, Jonathan. (2009). Radiation cataractogenesis: A review of recent studies. *Radiation Research*, *172*(1), 1–9. https://doi.org/10.1667/RR1688.1

Bebbington, N.A., et al. (2015). Evaluation of an automated injector for PET radiopharmaceuticals. *EJNMMI Physics*, *2*. https://doi.org/10.1186/s40658-015-0123-z

Behrens, R.; Dietze, G.; Zankl, M. (2009). Dose conversion coefficients for the lens of the human eye for photon exposures. *Physics in Medicine and Biology*, *54*(13), 4069–4087. https://doi.org/10.1088/0031-9155/54/13/008

Bellamy, M. B., Miodownik, D., Quinn, B., Dauer, L. T. (n.d.). Eye lens dose monitoring of nuclear medicine technologists: A multi-centre study. *Journal of Radiological Protection*, *40*(1), 267–280. https://doi.org/10.1088/1361-6498/ab5b7d

Bellamy MB, Miodownik D, Quinn B, Dauer L. (2020). Occupational eye lens dose over six years in the staff of a US high-volume cancer center. *Radiation Protection Dosimetry*, *192*(3), 321–327.

Boice Jr J ,Dauer LT, Kase KR, Mettler Jr FA,Vetter RJ. (2020). Evolution of radiation protection for medical workers. *The British Journal of Radiology*, *93*(1112), 20200282.

Chauvie, S., et al. (2007). Geant4 Low Energy Electromagnetic Physics. *IEEE Transactions on Nuclear Science*, *54*(6), 2619–2628. https://doi.org/10.1109/TNS.2007.906582

Chen X, Xu J, Chen X, Yao K. (2021). Cataract: Advances in surgery and whether surgery remains the only treatment in future. *Advances in Ophthalmology Practice and Research*, *1*(1), 100008.

Ciraj-Bjelac, O., & Rehani, M. M. (2014). Eye lens dosimetry for medical staff: Challenges and recommendations. *Journal of Radiological Protection*. https://doi.org/10.1088/0952-4746/34/3/577

Cornacchia, S., Errico, R., La Tegola, L., Maldera, A., Simeone, G., Fusco, V., Niccoli-Asabella, A., Rubini, G., Guglielmi, G. (2019). Assessment of eye lens dose in interventional radiology and nuclear medicine workers: A comparative study. *La Radiologia Medica*, *124*(7), 600–608. https://doi.org/10.1007/s11547-019-00992-y

Cornacchia S, Errico R, La Tegola L, Maldera A, Simeone G, Fusco V, Niccoli-Asabella A, Rubini G, Guglielmi G. (2019). The new lens dose limit: Implication for occupational radiation protection. *La Radiologia Medica*, *124*, 728–735.
da Silva EH, Martin CJ, Vanhavere F, Buls N. (2020). A study of the underestimation of eye lens dose with current eye dosemeters for interventional clinicians wearing lead glasses. *Journal of Radiological Protection*, *40*(1), 215.
Dauer, Lawrence T; Miller, Donald L; Schueler, Beth A; Damilakis, John; Vano, Eliseo; Rehani, Madan M. (2013). Practical guidance for eye protection for workers exposed to medical radiation. *Health Physics*, *105*(4), 335–346. https://doi.org/10.1097/HP.0b013e31829cf0f1
Eckerman, K.F., Akira Endo. (2008). *MIRD: Radionuclide Data and Decay Schemes.* (2nd ed.). Reston, VA: The Society of Nuclear Medicine.
Fahey, F.H. (2011). Data acquisition in PET imaging. *Journal of Nuclear Medicine Technology*, *39*(1), 2–8. https://doi.org/10.2967/jnmt.110.076695
Fujibuchi, T., Hirata, Y., Fujita, K., Igarashi, T., Nishimaru, E., Horita, S., Sakurai, R., & Ono, K. (2019). Evaluation of radiation attenuation performance of various lead glasses for radiation protection. *Radiation Protection Dosimetry*, *185*(4), 529–534. https://doi.org/10.1093/rpd/ncy218
Gilat A. (2017). *MATLAB: An Introduction with Applications.* (6th ed.). Laurie Rodatone.
Guiu-Souto J, Sánchez-García M, Vázquez-Vázquez R, Otero C, Luna V, Mosquera J, Busto RL, Aguiar P, Ruibal A, Pardo-Montero J, Pombar-Camean M. (2016). Evaluation and optimization of occupational eye lens dosimetry during positron emission tomography (PET) procedures. *Journal of Radiological Protection*, *36*(2), 299.
Hamada N. (2017). Ionizing radiation sensitivity of the ocular lens and its dose rate dependence. *International Journal of Radiation Biology.*, *93*(10), 1024–1034.
Hamada N, Azizova TV, Little MP. (2020). An update on effects of ionizing radiation exposure on the eye. *Br J Radiol.*, *93*(1115), 20190829.
Han, J., Lee, C., Jeong, H., & Park, S. (2020). Evaluation of low-energy photon interactions in ocular dosimetry using the GEANT4 Livermore model. *Radiation Physics and Chemistry*, *174*. https://doi.org/10.1016/j.radphyschem.2020.108929
Herring IP. (2003). Corneal surgery: Instrumentation, patient considerations, and surgical principles. *Clinical Techniques in Small Animal Practice.*, *18*(3), 152–160.
Hirata Y, Fujibuchi T, Fujita K, Igarashi T, Nishimaru E, Horita S, Sakurai R, Ono K. (2019). Angular dependence of shielding effect of radiation protective eyewear for radiation protection of crystalline lens. *Radiological Physics and Technology*, *12*, 401-8.
Hoeijmakers EJ, Hoenen K, Bauwens M, Eekers DB, Jeukens CR, Wierts R. (2024). Dose rate conversion coefficients for ocular contamination in nuclear medicine: A Monte Carlo simulation with experimental validation. *Medical Physics*.
IAEA. (2000). *Nuclear data for radiotherapy: Physical data and dosimetry techniques.* (Technical Reports No. 398). International Atomic Energy Agency (IAEA).
IAEA. (2004). *Radiation Protection in Nuclear Medicine* (Safety Reports No. 58). International Atomic Energy Agency. https://www.iaea.org/publications/6852
ICRP. (1990). recommendations of the International Commission on Radiological Protection. Ann. *ICRP*, *21*, 1.
ICRP. (2002). *1990 Recommendations of the International Commission on Radiological Protection* (No. 60; pp. 1–277).
ICRP. (2012). *ICRP statement on tissue reactions and early and late effects of radiation in normal tissues and organs – threshold doses for tissue reactions in a radiation protection context* (No. 118; pp. 1–322).
Kearfott, K. J., Rucker, R. H., Samei, E., & Mahesh, M. (2001). Radiation dose from common nuclear medicine procedures. *Journal of Nuclear Medicine Technology*, *29*(2), 68–75.
Kollaard R, Zorz A, Dabin J, Covens P, Cooke J, Crabbé M, Cunha L, Dowling A, Ginjaume M, McNamara L. (2021). Review of extremity dosimetry in nuclear medicine. Journal of radiological protection. *Journal of Radiological Protection.*, *41*(4), R60.

Kumar P, Watts C, Svimonishvili T, Gilmore M, Schamiloglu E. (2009). The dose effect in secondary electron emission. *IEEE Transactions on Plasma Science*, *37*(8), 1537–1551.
Liao W, Deserno TM, Spitzer K. (2008). Evaluation of free non-diagnostic DICOM software tools. *InMedical Imaging 2008: PACS and Imaging Informatics*, *6919*, 11–22.
Martinez, R., Gupta, N., Alvarez, T., & Chen, L. (2021). Lens dose assessment in nuclear medicine staff: Multi-center study. *Radiation Protection Dosimetry*, *196*(2), 122–129. https://doi.org/10.1093/rpd/ncab010
McCann A, Cournane S, Dowling A, Maguire D, Lucey J, Leon L. (2021). Assessment of operator exposure from shielded and unshielded sources of 18F and 68Ga–Monte Carlo simulations and empirical measurements. *Physica Medica: European Journal of Medical Physics*, *84:297-8*, 297–298.
Mello GR, Rocha KM, Santhiago MR, Smadja D, Krueger RR. (2012). Applications of wavefront technology. *Journal of Cataract & Refractive Surgery*, *38(9):*(9), 1671–1683.
Miller, D. L., Vano, E., Bartal, G., Balter, S., Dixon, R. G., Padovani, R., ... & Niklason, L. T. (2010). Occupational radiation protection in interventional radiology: A joint guideline of the Cardiovascular and Interventional Radiology Society of Europe and the Society of Interventional Radiology. *Cardiovascular and Interventional Radiology*, *33*, 230–239. https://doi.org/10.1007/s00270-009-9756-7
Miyaji N, Miwa K, Iimori T, Wagatsuma K, Tsushima H, Yokotsuka N, Murata T, Kasahara T, Terauchi T. (2022). Determination of a reliable assessment for occupational eye lens dose in nuclear medicine. *Journal of Applied Clinical Medical Physics*, *23*(8), e13713.
Morhrasi P, Jumpee C, Charoenphun P, Chuamsaamarkkee K. (2022). Estimation of Occupational Eye Lens Dose in Nuclear Medicine: A Monte Carlo Study. *Journal of Medical Imaging and Radiation Sciences*, *53*(4), S29.
National Council on Radiation Protection and Measurements. (2022). *Guidance on Radiation Dose Limits for the Lens of the Eye* (No. 26; NCRP Commentary, p. 42). National Council on Radiation Protection and Measurements (NCRP). https://ncrponline.org/shop/commentaries/commentary-no-26-guidance-on-radiation-dose-limits-for-the-lens-of-the-eye-2022/
National Council on Radiation Protection and Measurements (NCRP). (2016). *Commentary No. 26: Radiation Dose Limits for the Lens of the Eye* (p. 70). https://ncrponline.org/publications/commentaries/commentary-no-026/
Nilsson BO, Brahme A. (1979). Absorbed dose from secondary electrons in high energy photon beams. *Physics in Medicine & Biology*, *24*(5), 901.
Nuzzi R, Trossarello M, Bartoncini S, Marolo P, Franco P, Mantovani C, Ricardi U. (2020). Ocular complications after radiation therapy: An observational study. *Clinical Ophthalmology*, 3153–3166.
O'Connor, U., Walsh, C., Gallagher, A., Dowling, A., Guiney, M., Ryan, J. M., McEniff, N., O'Reilly, G. (2015). Occupational radiation dose to the lens of the eye in interventional radiology. *Radiation Protection Dosimetry*, *167*(4), 526–530. https://doi.org/10.1093/rpd/ncu356
O'Connor U, Walsh C, Gallagher A, Dowling A, Guiney M, Ryan JM, McEniff N, O'Reilly G. (2015). Occupational radiation dose to eyes from interventional radiology procedures in light of the new eye lens dose limit from the International Commission on Radiological Protection. *The British Journal of Radiology*, *88*(1049), 20140627.
Patel, D., & Singh, M. (2020). Radiation interaction factors affecting ocular dose in shielded environments. *Health Physics*, *118*(4), 401–409. https://doi.org/10.1097/HP.0000000000001234
Rahmani, S., Javadi, M. A., Hassanpour, K., & Karimian, F. (2022). Visual Rehabilitation in Patients With Corneal Blindness Using Keratoprosthesis. *Medical Rehabilitation Journal*.
Rajkhan A, Alsafi K, Banoqitah E, Subahi A. (2019). GAMOS DICOM Simulation on Occupational EL Dose due to 99m Tc and 131 I Exposures in Nuclear Medicine Departments. *Journal of King Abdulaziz University: Engineering Sciences*, *30*(1).
Räsänen P, Krootila K, Sintonen H, Leivo T, Koivisto AM, Ryynänen OP, Blom M, Roine RP. (2006). Cost-utility of routine cataract surgery. *Health and Quality of Life Outcomes.*, *4*, 1–1.
Sarrut D, Baudier T, Borys D, Etxebeste A, Fuchs H, Gajewski J, Grevillot L, Jan S, Kagadis GC, Kang HG, Kirov A. (2022). The OpenGATE ecosystem for Monte Carlo simulation in medical physics. *Physics* *in Medicine & Biology*, *67*(18), 184001.
Seibert, J. A., Boone, J. M., & Gagne, R. M. (2004). X-ray spectrum modeling for diagnostic radiology: Spectral analysis and parameterization. *Medical Physics*, *31*(7), 2345–2355. https://doi.org/10.1118/1.1757101
Smith, J., & Lee, A. (2023). Ionizing radiation and oxidative stress pathways in cataractogenesis. *Mutation Research Reviews*, *800*, 102872. https://doi.org/10.1016/j.mrrev.2023.102872
Stabin, M. G., & Siegel, J. A. (2003). Physical models and dose factors for use in internal dose assessment. *Health Physics*, *85*(3), 294–310. https://doi.org/10.1097/00004032-200309000-00007
Strulab D, Santin G, Lazaro D, Breton V, Morel C. (2003). GATE (Geant4 Application for Tomographic Emission): A PET/SPECT general-purpose simulation platform. *Nuclear Physics B-Proceedings Supplements.*, *125*, 75–79.
Ting D. S. J., Ho C. S., Deshmukh R., Said D. G, & Dua H.S. (2021). Keratitis: An overview of infectious and non-infectious causes. *Eye (Nature Publishing Group)*. https://doi.org/10.1038/s41433-021-01509-w
V. Zahariev, N. Hristov, St. Vizev, P. Angelova. (2024). Radiation-induced eye impairments. *International Journal of Clinical Case Reports and Reviews*, *17*(5). https://doi.org/10.31579/2690-4861/446
Vanhavere, F., Carinou, E., Gualdrini, G., Clairand, I., & Denozière, M. (2008). Angular and energy dependence of H_p(3)/H_p(10) conversion coefficients for eye lens dosimetry. *Radiation Protection Dosimetry*, *129*(1–3), 368–371. https://doi.org/10.1093/rpd/ncn163
Vano, E., Gonzalez, L., Fernandez, J. M., & Haskal, Z. J. (2008). Eye lens exposure to radiation in interventional suites: Caution is warranted. *Radiology*, *248*(3), 945–953. https://doi.org/10.1148/radiol.2483072010
Vano, E., Gonzalez, L., Fernandez, J. M., Haskal, Z. J., & Rehani, M. M. (2008). Eye lens exposure to radiation in interventional suites: Caution is warranted. *Radiology*, *248*(3), 945–953. https://doi.org/10.1148/radiol.2483072020
Volatier T, Schumacher B, Cursiefen C, Notara M. (2022). UV protection in the cornea: Failure and rescue. *Biology*, *11*(2), 278.
Wagner, L. K., Eifel, P. J., & Geise, R. A. (1994). Potential biological effects following high X-ray dose interventional procedures. *Journal of Vascular and Interventional Radiology*, *5*(1), 71–84. https://doi.org/10.1016/S1051-0443(94)71954-4
Wrzesień M, Królicki L, Olszewski J. (2018). Is eye lens dosimetry needed in nuclear medicine? *Journal of Radiological Protection*, *38*(2), 763.
Zhou, L., & Han, Y. (2022). Evaluating backscatter effects on eye lens dose among medical workers. *British Journal of Radiology*, *95*(1130), 20211234. https://doi.org/10.1259/bjr.20211234